\documentclass{article}
\usepackage{graphicx}
\usepackage{epsfig}
\usepackage{amsfonts}
\usepackage{amsmath, amssymb, array}
\usepackage{graphics,graphicx,float,mfpic}
\usepackage{epsfig}
\usepackage{color}
\newcommand{\be}{\begin{equation}}
\newcommand{\ee}{\end{equation}}
\newcommand{\bea}{\begin{eqnarray}}
\newcommand{\eea}{\end{eqnarray}}
\newcommand{\bmat}{\begin{pmatrix}}
\newcommand{\emat}{\end{pmatrix}}

\begin{document}
\title{On the stability of AdS black strings } 
 
\author{{\large Yves Brihaye,}$^{\dagger}$
{\large Terence Delsate }$^{\dagger}$
and {\large Eugen Radu}$^{\ddagger }$ \\ \\
$^{\dagger}${\small Physique-Math\'ematique, Universite de
Mons-Hainaut, Mons, Belgium}\\
$^{\ddagger}${\small Laboratoire de Math\'ematiques et Physique Th\'eorique,
Universit\'e Fran\c{c}ois-Rabelais, Tours, France}
  }

\maketitle

\begin{abstract}
We explore via linearized perturbation theory 
the Gregory-Laflamme instability of the black string solutions of Einstein's 
equations with negative cosmological constant
recently discussed in literature. 
Our results indicate that the black strings whose
 conformal infinity is the product of time and $S^{d-3}\times S^1$
are stable for large enough values of the event horizon radius.
All topological black strings are also classically stable. 
We argue that this provides an explicit realization of the Gubser-Mitra conjecture.
\end{abstract}

%%%%%%%%%%%%%%%%%%%%%%%%%%%%%%%%%%%%%%%%%%%%%%%%%%%%%%%%%%%%%%%%%%%%
\section{Introduction}
%%%%%%%%%%%%%%%%%%%%%%%%%%%%%%%%%%%%%%%%%%%%%%%%%%%%%%%%%%%%%%%%%%%%
In 1993, Gregory and Laflamme (GL) made the surprising discovery that the extended black objects 
appearing in higher dimensions are classically unstable \cite{Gregory:1993vy}.
The simplest extended black object in general relativity
is found by trivially extending to $d$-spacetime dimensions  
the Schwarzschild black hole solution in $d-1$ dimensions and
 corresponds to a uniform black string  (UBS)
with horizon topology $S^{d-3}\times S^1$. 
The results in \cite{Gregory:1993vy} showed that
this configuration is classically unstable below a critical value of the mass. 
Following this discovery, a branch of nonuniform black string (NUBS)
solutions was found perturbatively from the critical
GL string  \cite{Gubser:2001ac}, \cite{Wiseman:2002zc}, \cite{Sorkin:2004qq}. 
This nonuniform branch was subsequently numerically extended into the full nonlinear regime
in  \cite{Wiseman:2002zc}, \cite{Kleihaus:2006ee}, \cite{Sorkin:2006wp}
(see \cite{Kol:2004ww},  \cite{Harmark:2007md} 
for reviews of this topic).

Most of the work on the stability and phases of black strings have been performed so far 
assuming a vanishing cosmological constant.
However, recently there has been some interest in finding black string solutions
with a cosmological constant $\Lambda$.
The anti-de Sitter (AdS) natural counterparts of the Schwarzschild black string 
have been considered for the first time in 
 \cite{Copsey:2006br}, for the  $d=5$ case.
These 
UBS solutions 
are very different from the warped  AdS black 
strings as discussed for instance in \cite{Chamblin:1999by}  (see also \cite{Hirayama:2001bi}).

For example, they have no 
dependence on the `compact' extra dimension, 
and their conformal boundary is the product of 
time and $S^{2}\times S^1$.
Higher 
dimensional  $d>5$ black strings have been constructed 
in \cite{ Mann:2006yi},  
configurations with an event horizon topology 
$H^{d-3}\times S^1$ being considered as well. 
As argued in \cite{Copsey:2006br, Mann:2006yi}, these solutions
provide the gravity dual of a field theory  on
a $S^{d-3}\times S^1\times S^1$ (or $H^{d-3}\times S^1\times S^1$)
background.
Various other examples of AdS black strings, including solutions with 
matter fields, can be found in 
\cite{Brihaye:2007vm}, \cite{Brihaye:2007ju}, \cite{Bernamonti:2007bu}.
Black string with a positive cosmological constant
have been constructed  in  \cite{Brihaye:2006hsa}.
Different from the $\Lambda=0$ limit,  the AdS black 
string solutions  with an event horizon 
topology $S^{d-3}\times S^1$ have a nontrivial, 
globally regular limit with zero event 
horizon radius  \cite{ Mann:2006yi}.

However, the issue of the classical stability of the AdS black strings 
has not yet been addressed in the literature.
On general grounds, one expects a richer structure in this case, since the cosmological constant
introduces another scale in the theory.
This type of solutions also provides a new laboratory
to test the Gubser-Mitra (GM) conjecture \cite{Gubser:2000ec}, 
that correlates
the dynamical and thermodynamical stability 
for systems with translational symmetry and infinite extent.
In this conjecture, the appearance of a negative specific heat is related to the onset of a classical instability.
This conjecture has passed a large number of tests \cite{Harmark:2007md}, \cite{Hirayama:2001bi},
but it is also known to fail in certain cases
\cite{Friess:2005zp}.  
Since thermodynamical  stability of the  AdS black strings is possible for $\Lambda<0$ \cite{ Mann:2006yi}, one expects 
that some of these objects
do not present the GL instability. 

The paper is structured as follows:
we begin with a review of the general properties of the 
AdS UBS solutions. 
In  Section 3, we consider  the issue of GL instability.
Expanding around the UBS and solving the eigenvalue problem numerically,
our results indicate that the GL
instability persists up to a critical configuration only.
 The final Section contains a discussion of the  
GM conjecture for AdS black strings together with our conclusions. 

%%%%%%%%%%%%%%%%%%%%%%%%%%%%%%%%%%%%%%%%%%%%%%%%%%%%%%%%%%%%%%%%%%%%%%%%%%%%%%%%%%%%%%%%%%%%%%%%
\section{AdS uniform black strings}
%%%%%%%%%%%%%%%%%%%%%%%%%%%%%%%%%%%%%%%%%%%%%%%%%%%%%%%%%%%%%%%%%%    
These solutions of the Einstein equations with a negative
cosmological constant  $\Lambda=-(d-1)(d-2)/(2 \ell^2)$ 
have been found for a metric ansatz with three unknown functions \cite{ Mann:2006yi}
\begin{eqnarray}
\label{metricUBS} 
ds^2 = -b(r) dt^2 +\left(\frac{dr^2}{f(r)} + a(r)dz^2\right) + r^2 d\Sigma^2_\kappa~,
\end{eqnarray} 
where the $(d{-}3)$--dimensional metric $d\Sigma^2_{\kappa,d-3}$ is 
\begin{equation}
d\Sigma^2_{\kappa,d-3} =\left\{ \begin{array}{ll}
\vphantom{\sum_{i=1}^{d-3}}
 d\Omega^2_{d-3}& {\rm for}\; \kappa = +1\\
\sum_{i=1}^{d-3} dx_i^2&{\rm for}\; \kappa = 0 \\
\vphantom{\sum_{i=1}^{d-3}}
 d\Xi^2_{d-3} &{\rm for}\; \kappa = -1\ ,
\end{array} \right.
\end{equation}
with $d\Omega^2_{d-3}$  the unit metric on $S^{d-3}$; by $H^{d-3}$ 
we will understand the $(d{-}3)$--dimensional hyperbolic space, whose unit 
metric  $d\Xi^2_{d-3}$ can be obtained by analytic continuation of 
that on $S^{d-3}$.  
As usual with black strings, the coordinate along the compact direction is denoted by $z$ and its
asymptotic length is $L$.

Unfortunately, for  $\Lambda \neq 0$,
the equations satisfied by the metric functions $a,b$ and $f$
can  be solved only numerically (these equations are presented in \cite{ Mann:2006yi}). 
 However, one can write an approximate form of the solutions
 valid near the event horizon or for large values of the radial coordinate  $r$.
The event horizon is taken at constant  $r=r_h$, where one finds an expansion
in terms of two parameters $a_h$, $b_1$ 
\begin{eqnarray} 
\label{eh}
&a(r)=
a_h
%+a_1(r-r_h)
+O(r-r_h),
~
b(r)=b_1(r-r_h) 
+O(r-r_h)^2,
~
f(r)=\bar f_1(r-r_h)
+O(r-r_h)^2,
\end{eqnarray}
with $\bar f_1= \big((d-1)r_h^2+k(d-4)\ell^2\big)/(r_h \ell^2)$.
 The condition for a regular 
event horizon is $\bar f_1>0$, $b_1>0$. In the $\kappa=-1$ case, this 
implies the existence of a minimal value of $r_h$.
However, for any value of $\kappa$, there is no upper bound on  
the event horizon radius.
In Refs. \cite{Copsey:2006br, Mann:2006yi} arguments
 for the existence
 of a nontrivial globally regular solutions with $r_h=0$ in the case $\kappa=1$
 are presented.
Also,
for $\kappa=0$, the Einstein equations admit the exact solution $a=r^2$, 
$f=1/b=-2M/r^{d-3}+r^2/\ell^2$, which  appears to be unique, corresponding to the 
known planar topological black hole with a periodic $z$-direction.

The expression of the solution near the boundary at infinity  
depends on two constants $c_t$ and $c_z$, which
fix the UBS mass and tension.
For even $d$, the solution
admits at large $r$  a power series expansion of the form:
\begin{eqnarray} 
\nonumber
&a(r)=\frac{r^2}{\ell^2}+\sum_{j=0}^{(d-4)/2}a_j(\frac{\ell}{r})^{2j}
+c_z(\frac{\ell}{r})^{d-3}+O(1/r^{d-2}),~~
b(r)=\frac{r^2}{\ell^2}+\sum_{j=0}^{(d-4)/2}a_j(\frac{\ell}{r})^{2j}
+c_t(\frac{\ell}{r})^{d-3}+O(1/r^{d-2}),
\end{eqnarray}  
\begin{eqnarray} 
\label{even-inf}
&f(r)=\frac{r^2}{\ell^2}+\sum_{j=0}^{(d-4)/2}f_j(\frac{\ell}{r})^{2j}
+(c_z+c_t)(\frac{\ell}{r})^{d-3}+O(1/r^{d-2}),
\end{eqnarray}   
where $a_j,~f_j$ are constants depending on the index
$\kappa$ and of the spacetime dimension only.
 Their expression can be found in   \cite{ Mann:2006yi}.
The corresponding expansion for odd values of the spacetime dimension contains log terms
and is given by:
\begin{eqnarray}
\label{odd-inf}
a(r)=\frac{r^2}{\ell^2}+\sum_{j=0}^{(d-5)/2}a_j(\frac{\ell}{r})^{2j}
+\zeta\log(\frac {r}{\ell}) (\frac{\ell}{r})^{d-3}
+c_z(\frac{\ell}{r})^{d-3}+O(\frac{\log r}{r^{d-1}}),
\end{eqnarray} 
\begin{eqnarray}
\nonumber
b(r)&=&\frac{r^2}{\ell^2}+\sum_{j=0}^{(d-5)/2}a_j(\frac{\ell}{r})^{2j}
+\zeta\log (\frac {r}{\ell}) (\frac{\ell}{r})^{d-3}
+c_t(\frac{\ell}{r})^{d-3}+O(\frac{\log r}{r^{d-1}}),
\\
\nonumber
f(r)&=&\frac{r^2}{\ell^2}+\sum_{j=0}^{(d-5)/2}f_j(\frac{\ell}{r})^{2j}
+2\zeta\log (\frac {r}{\ell}) (\frac{\ell}{r})^{d-3}
+(c_z+c_t+c_0)(\frac{\ell}{r})^{d-3}+O(\frac{\log r}{r^{d-1}}),
\end{eqnarray}   
where $c_0,~\zeta$ are constants (supplementing the $a_j,f_j$ mentionned above)
 depending on
$\kappa$, $d$ \cite{ Mann:2006yi}, .
 
The mass 
 of the UBS solutions as evaluated in  \cite{ Mann:2006yi} by using the standard
counterterm method  is (in this paper we take $G_d=1$)
\begin{eqnarray}
\label{MT} 
M&=&\frac{\ell^{d-4}}{16\pi   
}\big[c_z-(d-2)c_t\big]LV_{\kappa,d-3}+M_c^{(\kappa,d)}~, 
\end{eqnarray}
where $V_{\kappa,d-3}$ is the total area of the angular sector.
$M_c^{(\kappa,d)}$  is a  Casimir-like term 
which appears for an odd spacetime dimension only, $M_c^{(\kappa,d)} = \frac{\ell^{d-4}}{16\pi 
}V_{\kappa,d-3}L\left(\frac{1}{12}\delta_{d,5}
-\frac{333}{3200}\delta_{d,7}+\dots\right)$.
%(these solutions have also a nonzero  tension, which has a rather similar expression).
The Hawking temperature and the entropy of the UBS are given by 
\begin{eqnarray}
T_H=\frac{1}{4\pi}\sqrt{\frac{b_1}{r_h\ell^2}\big[(d-1)r_h^2+\kappa(d-4)\ell^2\big]},~~
S=\frac{1}{4 }r_h^{d-3}V_{\kappa,d-3}L\sqrt{a_h}.
\label{TS}
\end{eqnarray} 

The thermodynamics of the vacuum AdS UBSs has been discussed to some 
extend in \cite{ Mann:2006yi},
where it was found that they follow  the pattern of the 
corresponding Schwarzschild-AdS$_d$ (SAdS$_d$) black hole solutions.
In the more interesting $\kappa=1$ case,
the temperature of the  
black string solutions is bounded from below.
At higher temperatures, above a critical value, 
there exist two bulk solutions that correspond to the so-called small 
(unstable) and large (stable) black string solutions. 
Concerning local  thermodynamic stability, 
in the case of a canonical ensemble  with a fixed value of the length of 
the extra-dimension $L$,   the response function whose
 sign determines the thermodynamic stability is the heat capacity
 $C=T_H ( {\partial S}/{\partial T_H} )_{L}$. 
The results in \cite{ Mann:2006yi} indicate that the
 small UBSs with $\kappa = 1$  have negative
specific heat (they are thermodynamically unstable) but large size black strings have
positive specific heat (and they are stable).
The topological black strings ($\kappa=0,-1$) are thermodynamically stable.

We close this brief review of the UBS properties, by noting that the Einstein 
equations  are left invariant by the transformation  
 $
r \to \bar{r}= \xi r,~~\ell \to \bar{\ell}= \xi \ell.
$
Therefore,  one may 
generate in this way a family of  new vacuum solutions, which are usually termed 
``copies of solutions`` \cite{Harmark:2003eg}. 
The new solutions have the 
same length in the extra-dimension, the physical quantities vary according to
\begin{eqnarray}
\label{transf2} 
 \bar{r}_h=\xi r_h,~\bar{\Lambda}=\Lambda/\xi^2 ,~~
 \bar{T}_H=T_H/\xi ,~~
 \bar{M}=\xi^{d-4} M ,~~\bar{S}=\xi^{d-3} S .
\end{eqnarray}
As a consequence, given the full spectrum of solutions for a given value of $\Lambda$ (with
$r_{min}<r_h<\infty$), one may find the corresponding branches for any value of 
$\Lambda<0$. 
Equivalently, one can cover the full UBS spectrum by keeping fixed the event horizon radius and 
 varying the value of the cosmological constant $0<\left|  \Lambda\right|<\infty$.

The problem is thus characterized by two dimensionless parameters
%\begin{eqnarray}
%\label{ls} 
$ \mu_1= {M}/{L^{d-3}},~~\mu_2= {L}/{\ell}~$.
%\end{eqnarray} 
The limit $\mu_2\to 0$ corresponds to black string solutions in a Kaluza-Klein theory.

%%%%%%%%%%%%%%%%%%%%%%%%%%%%%%%%%%%%%%%%%%%%%%%%%%%%%%%%%%%%%%%%%%
\section{The Gregory--Laflamme instablity}
%%%%%%%%%%%%%%%%%%%%%%%%%%%%%%%%%%%%%%%%%%%%%%%%%%%%%%%%%%%%%%%%%%
The GL mode of the AdS black strings is found in this paper  by using the 
same approach as in \cite{Gubser:2001ac}.
A convenient ansatz for the static NUBS solutions of the Einstein  equations
with negative cosmological constant is  
\begin{eqnarray}
\label{metric1} 
ds^2 = -b(r) e^{2A(r,z)}dt^2 + e^{2B(r,z)}\left(\frac{dr^2}{f(r)} 
+ a(r)dz^2\right) + r^2e^{2C(r,z)}d\Sigma^2_\kappa~.
\end{eqnarray}
The UBS limit discussed above corresponds to $ A=B=C=0$.
This metric form generalizes for the AdS case 
the usual $\Lambda=0$ NUBS ansatz used e.g. in \cite{Wiseman:2002zc}, \cite{Kleihaus:2006ee}, \cite{Sorkin:2006wp}, 
which  is recovered for $\kappa = 1$
and 
$ b=1/f=1- (r_0/r)^{d-4},~a=1$.

Following the standard approach, 
we perform an expansion around  the UBS of the form
\bea
\label{n11}
A(z,r) = \lambda A_1(r)\cos(kz) + O(\lambda^2),~~
B(z,r) = \lambda B_1(r)\cos(kz) + O(\lambda^2),~~
C(z,r) = \lambda C_1(r)\cos(kz) + O(\lambda^2),
\eea
with $\lambda$ a small parameter. 
The ordinary differential equations for $A_1,B_1,C_1$ 
are found by plugging (\ref{n11}) into the Einstein equations and retaining the first order in
$\lambda$.
Following \cite{Gubser:2001ac}, we express the function $B_1$ 
in terms of  $A_1,~C_1$ and their first derivatives. This
 leads to a system of two differential equations:
\be
A_1''= \alpha_1 A_1 + \alpha_2 A_1' + \alpha_3 C_1 + \alpha_4 C_1',~~
C_1'' = \varphi_1 A_1 + \varphi_2 A_1' + \varphi_3 C_1 + \varphi_4 C_1',
\label{eqF1}
\ee
where  
\bea
\alpha_1 &=& \frac{2b\left( \left( -3 + d \right) k^2l^2 + \left( r - dr \right) a' \right)  + 
    r\left( k^2 \ell^2 + 2\left( -1 + d \right) a \right) b'}{\ell^2af
    \left( 2\left( -3 + d \right) b + rb' \right) },
\nonumber
\\
\label{eqs-p}
\alpha_2 &=&-\frac{1}{r} - \frac{b'}{2b} - 
  \frac{1}{\ell^2
     rf}\left[\left( -1 + d \right) r^2 + \left( -4 + d \right) \ell^2\kappa  - 
     \frac{4\left( -1 + d \right) r^2b}{2\left( -3 + d \right) b + rb'}\right],
 \\    
     \alpha_3 &=&\frac{2\left( -3 + d \right) \left( -1 + d \right) b\left( 2a - ra' \right) }
  {\ell^2af\left( 2\left( -3 + d \right) b + rb' \right) },~~
\alpha_4 =-\frac{\  \left( -3 + d \right) b'   }{2b} + 
  \frac{4\left( -3 + d \right) \left( -1 + d \right) rb}
   {\ell^2 f\left( 2\left( -3 + d \right) b + rb' \right) },
\nonumber
\\
\nonumber
\varphi_1 &=&\frac{2\left( \left( -1 + d \right) r^2 + \left( -4 + d \right) \ell^2\kappa  \right) 
    \left( -  b a'  + a b' \right) }{\ell^2 r a f
    \left( 2\left( -3 + d \right) b + r b' \right) },~~~
\varphi_2 =\frac{1}{r}\left[-1 + \frac{4 b\left( \left( -1 + d \right) r^2 + \left( -4 + d \right) \ell^2\kappa  \right)
         }{\ell^2 f\left( 2\left( -3 + d \right) b + rb' \right) }\right],
 \\        
         \nonumber
\varphi_3 &=&\frac{2b\left( -3 + d \right) \left( k^2 \ell^2 r + 2\left( -1 + d \right) ra - 
       \left( \left( -1 + d \right) r^2 + \left( -4 + d \right) \ell^2\kappa  \right) a'
       \right)  + \ell^2\left( k^2r^2 - 2\left( -4 + d \right) \kappa a \right) b'}
    {\ell^2 raf\left( 2\left( -3 + d \right) b + rb' \right) },
\nonumber
\\
\varphi_4 &=&  \frac{2b\left( -3 + d \right) \left( k^2 \ell^2r + 2\left( -1 + d \right) ra - 
       \left( \left( -1 + d \right) r^2 + \left( -4 + d \right) \ell^2\kappa  \right) a'
       \right)  + \ell^2\left( k^2r^2 - 2\left( -4 + d \right) \kappa a \right) b'}
    {\ell^2raf\left( 2\left( -3 + d \right) b + rb' \right) }.
\nonumber
\end{eqnarray}
 This eigenvalue problem for the wavenumber $k=2\pi/L$ was solved numerically
with suitable boundary conditions.
The solutions of (\ref{eqs-p}) were constructed in a 
systematic way  for $5 \le d \le 8$ and $\kappa=0,\pm 1$;
a number of solutions with $d=9,10,11$ have been also considered. 
To integrate the equations, we used the differential
equation solver COLSYS which involves a Newton-Raphson method
\cite{COLSYS}.
At $r=r_h$ the regularity of the solution
imposes specific relations between $A_1,C_1$ and 
their first derivatives; these relations are very long and we prefer
 not to include them  here.
Also, the perturbation has to vanish at $r \to \infty$, i.e.
 ${\rm lim}_{r\to \infty} A_1,C_1 = 0$. 
 
 Given  the linearity of the equations, there is also a freedom in the
 choice of the
normalisation of the solutions.
Solutions obeying the appropriate boundary conditions exist only for specific values
of the spectral parameter $k^2$. 
In practice, 
 we first computed the background functions $a,b$ and $f$, and then solve the equations (\ref{eqF1})
supplemented by the equation $d k^2/ dr = 0$.
The supplementary boundary condition allowed by this extra equation
can then be imposed in order to eliminate the arbitrariness of the
normalisation (we choose $C_1(r_h)=1$) and leads to a well posed
problem.
Once given a starting profile
sufficiently close to the solution,
the numerical solver is able to construct with a very good accuracy both 
the perturbation and the corresponding  value of $k^2$ for a given set
of the parameters ($\kappa, d, r_h, l^2$).
A positive value for $k^2$ indicates the presence of an unstable mode of the background UBS configuration.

Starting with the more familiar solutions with an event horizon topology
$S^{d-3}\times S^1$,  we noticed the existence, for a fixed value of $r_h$,
of a critical value of the cosmological constant $\Lambda_c$, beyond which the 
AdS black strings become stable. 
The value of $\Lambda_c$ is dimension dependent; one finds e.g.  $\Lambda_c(d=5)\simeq -1.27$, $\Lambda_c(d=6)\simeq -3.52$
(the event horizon radius here is   $r_h=1$). 
However, all solutions with $ \Lambda $ between zero and $\Lambda_c$ are classical unstable.
The precise value of the  wave-number $k$ depends on $d$ and $\Lambda$.

Taking instead a fixed value of the cosmological constant and varying the event horizon radius,
this translates into the existence
of a critical $r_h^c$ above which the solutions are stable (note that the product $k \ell$ 
is invariant under the rescaling of the radial coordinate).
These results are presented in Figure 1.
One can also see  that the quantity $k^2$ is positive and diverges 
as $r_h \to 0$. This demonstrates that
 AdS globally regular 
  %
  %
%%%%%%%%%%%%%%%%%%%%%%% Figure 1 %%%%%%%%%%%%%%%%%%%%%%%%%%
\newpage
\setlength{\unitlength}{1cm}
\begin{picture}(18,7)
\centering
\put(2,0.0){\epsfig{file=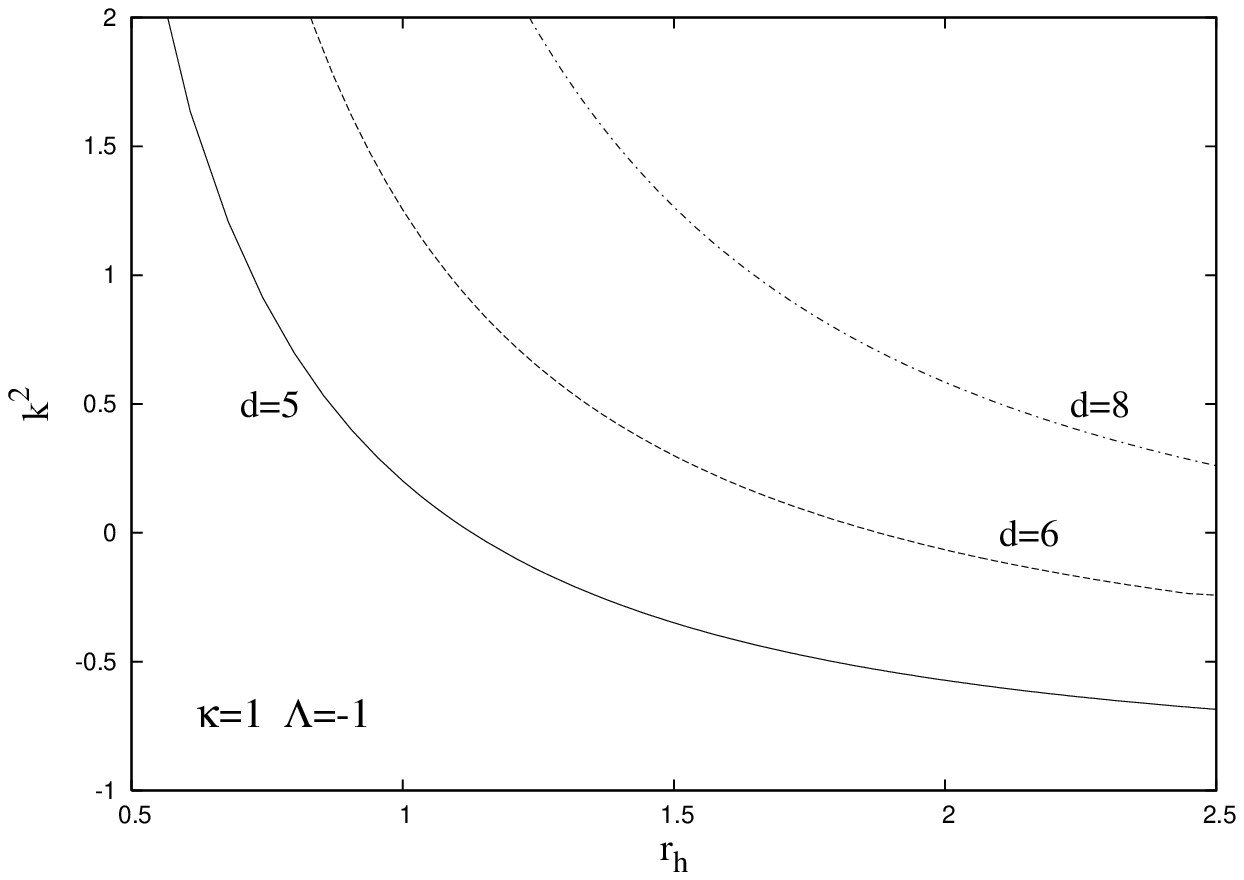,width=13cm}}
\end{picture} 
{\small {\bf Figure 1.}
The  square of the
critical wave-number $k^2$ is plotted 
as a function of the horizon radius $r_h$ for black string solutions with an event horizon topology
$S^{d-3}\times S^1$ and several spacetime dimensions. 
 }
\\
\\
%%%%%%%%%%%%%%%%%%%%%%% Figure 2 %%%%%%%%%%%%%%%%%%%%%%%%%%
\setlength{\unitlength}{1cm}
\begin{picture}(18,7)
\centering
\put(2.97,-2.20){\epsfig{file=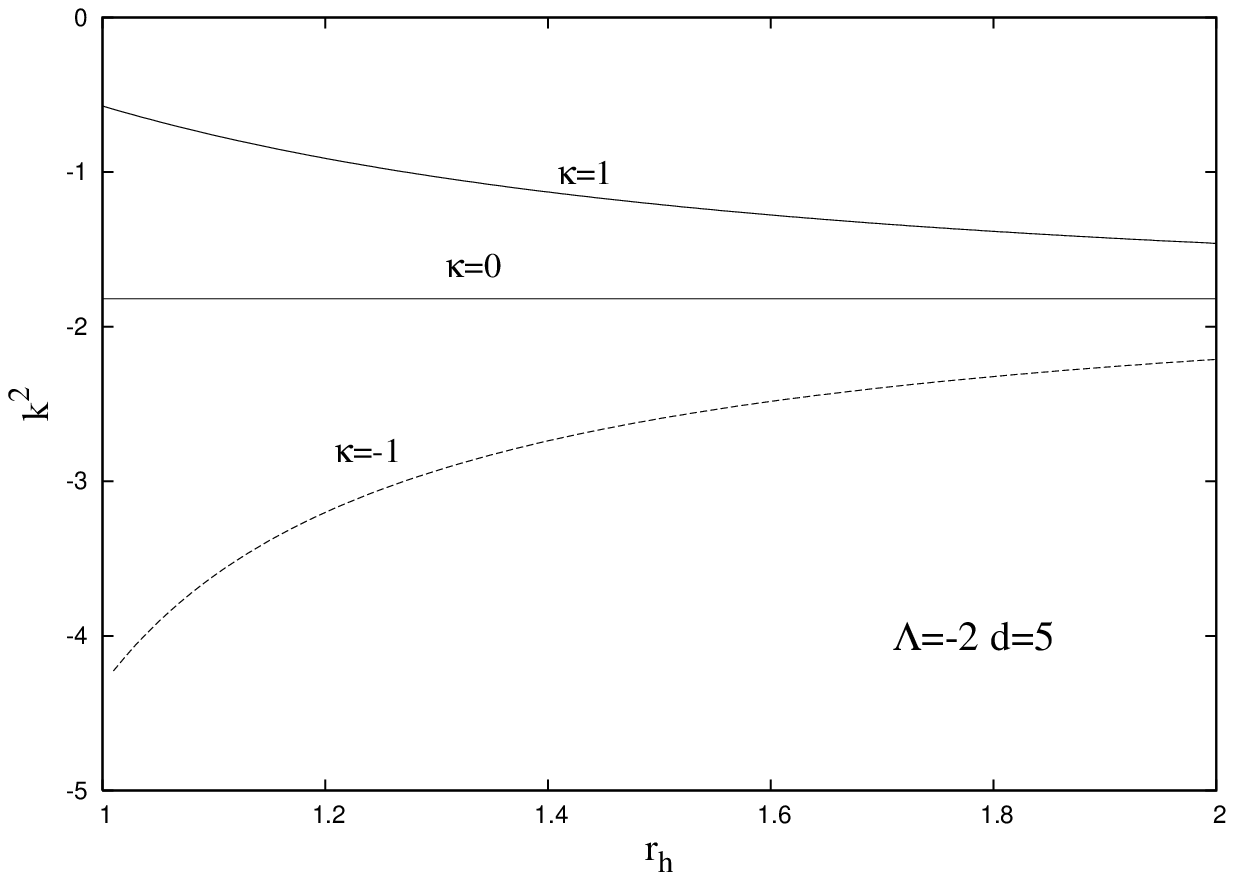,width=12.6cm}}
\end{picture} 
\\
\\
\\
\\
\\
\\
{\small {\bf Figure 2.}
The same as Figure 1 for $\Lambda=-2,\ d=5$ black string solutions  with $\kappa=-1,\ 0,\ 1$. }
\\
\\
UBS solution is classically unstable.

The picture for topological black strings ($\kappa=0,-1$) is very different.
Our findings indicate that the topological black string solutions 
do not present an instability within the ansatz
(\ref{metric1}), (\ref{n11}). The numerics indeed 
gives $k^2<0$ for all values of ($r_h,~\Lambda$) that we have considered.
The dependance of the eigenvalue $k^2$ on the horizon radius $r_h$ for the
three possible topologies (in the allowed region for $\kappa=-1$) 
is reported in Figure 2 for $d=5$ black strings with $\Lambda=-2$.
The value corresponding to $\kappa=0$ is $r_h$-independent and turns out to constitute a lower (resp. upper) bound for the case 
$\kappa=1$ (resp. $\kappa=-1$).

 %%%%%%%%%%%%%%%%%%
 %%%%%%%%%%%%%%%%%%%%%%% Figure 3 %%%%%%%%%%%%%%%%%%%%%%%%%%
\newpage
\setlength{\unitlength}{1cm}
\begin{picture}(18,7)
\centering
\put(2,0.0){\epsfig{file=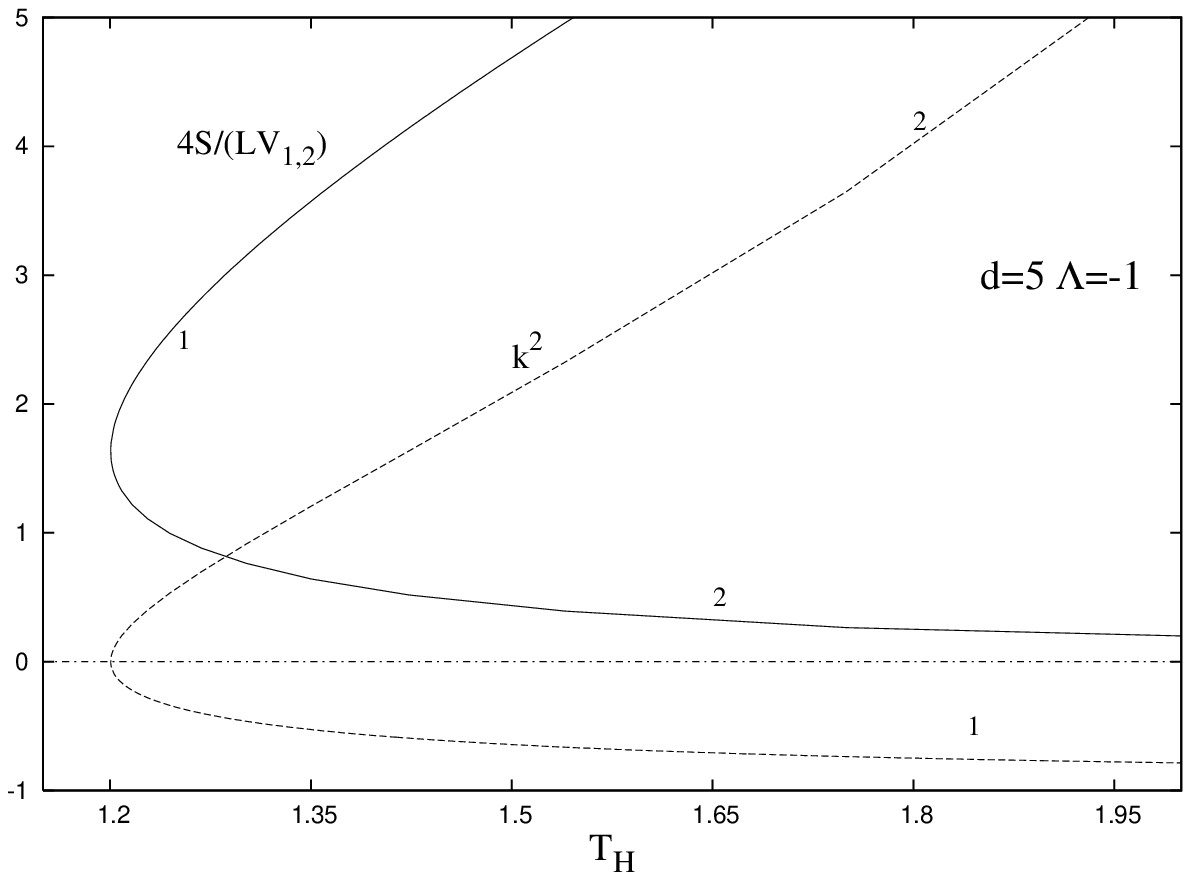,width=13cm}}
\end{picture}
\begin{picture}(19,9.)
\centering
\put(2.5,0.0){\epsfig{file=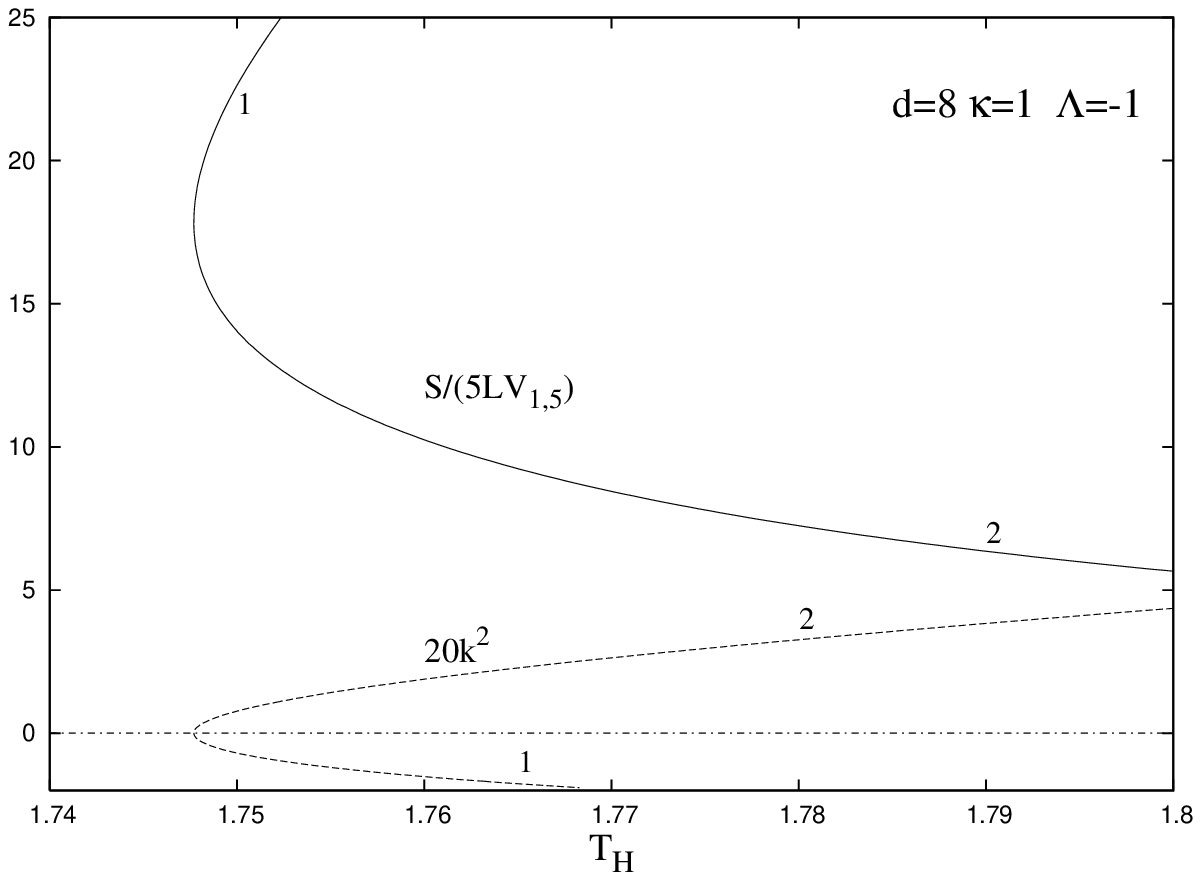,width=13cm}}
\end{picture}
\\
\\
{\small {\bf Figure 3.}
The entropy per unit surface $S/(4LV_{\kappa,d-3})$ and the square of the
critical wave-number $k^2$ are  plotted as a function of the Hawking temperature $T_H$ 
for black string solutions with an event horizon topology
$S^{d-3}\times S^1$ in $d=5$ and $d=8$ spacetime dimensions  
(for a better visualisation, both $S$ and $k^2$ are  multiplied with suitable factors).}
\\
\\
It is also interesting to notice that $k^2$ is well fitted by a function of the form
 \begin{equation}
  k^2 \simeq \alpha + \frac{\beta}{r_h^c}~;
 \end{equation}
  the best fit for the parameters $a,b,c$ gives
  $\alpha\simeq-1.85, \beta\approx 1.28, c\simeq 1.68$ for $\kappa = 1$
 and  $\alpha\simeq -2.02, \beta\simeq -2.22, c\simeq 3.4$ for $\kappa =-1$ and
 $\alpha= -1.82,$ $\beta=0$ for $\kappa = 0$ (with $\Lambda = -2$). 
 
%%%%%%%%%%%%%%%%%%%%%%%%%%%%%%%%%%%%%%%%%%%%%%%%%%%%%%%%%%%%%%%%%%
\section{Further remarks}
%%%%%%%%%%%%%%%%%%%%%%%%%%%%%%%%%%%%%%%%%%%%%%%%%%%%%%%%%%%%%%%% 
On general grounds, one expects at least the small AdS black strings 
to present a GL-type instability. This was confirmed by our numerical results 
discussed in the previous section
for   black strings with an horizon topology $S^{d-3}\times S^1$. 
Moreover, we have found that all topological black strings are stable 
with respect
to  perturbations of the form (\ref{n11}) emphasized in this paper.

 We expect the results of this paper to be relevant for the issue of AdS black rings (yet to be found).
 As observed in \cite{ Mann:2006yi}, the heuristic construction of black
rings (see $e.g.$ \cite{Emparan:2007wm}) applies also for $\Lambda<0$, and one expects an AdS black ring to approach
the UBS black string solution in the limit where the radius of the ring circle grows very large.
 Thus, in a similar way to the 
asymptotically flat case \cite{Elvang:2006dd}, some of the AdS black ring solutions will also present a GL-type instability.

The results in the previous Section provide also an explicit realization of the GM conjecture.
This is  illustrated in Figure 3, where we plot 
the wavelength $k$ and the entropy $S$ as a function of temperature $T_H$ for
$\kappa=1$ black strings with a fixed value of $\Lambda$, in $d=5,~8$ 
spacetime dimensions.
One can see in particular that the solutions with $k^2<0$
have also a positive specific heat (this branch has the index $1$ in that plot), 
while
the entropy decreases with $T_H$ along the branch with $k^2>0$ 
(index $2$ in the Figure 3).
In agreement with the GM conjecture,  
the critical temperature where $k^2$ crosses zero coincides 
(within the numerical accuracy) with
the temperature at the turning point of the UBS branches in a $(S-T_H)$ diagram 
($e.g.$  $T_H^c\simeq 1.2005$ for $d=5$ and $T_H^c\simeq 1.7477$ for $d=8$).
A similar picture has been found for other spacetime dimensions.
When considering instead topological black strings, the result  $k^2<0$
is, indeed, what we expect
since these solutions have always a positive specific heat \cite{ Mann:2006yi}.

In principle, following \cite{Gregory:1993vy},
one can get an estimation of the critical wave-number $k$ by equating 
the entropy of the UBS
with that of SAdS black hole in $d-$dimensions with the same mass and the same topological parameter $\kappa$.
 The  relation between the entropy 
and the mass of a  SAdS$_d$ black hole is  
\begin{eqnarray}
\label{1rel1}
\kappa-\frac{M}{\frac{(d-2)V_{\kappa,d-2}}{16\pi}}\frac{1}{(4S/V_{\kappa,d-2})^{(d-3)/(d-2)}}
+\left(\frac{4S}{V_{\kappa,d-2}}\right)^{\frac{2}{d-2}}\frac{1}{\ell^2}=0,
\end{eqnarray} 
(here we ignore the Casimir-like mass terms which appear for an odd $d$).
For a  UBS solution with the same values of mass and entropy, one can use (\ref{MT}), (\ref{TS})  to write $
S= {V_{\kappa,d-3} L \ell^{d-3}} n_1/{4}
$, 
$
M= {V_{\kappa,d-3} L \ell^{d-4}} n_2/{(16\pi)},
$
with the numerical coefficients    
$
n_1=\sqrt{a_h}( {r_h }/{\ell })^{d-3},~~n_2=c_z-(d-2)c_t.
 $ 
This yields the following equation for the critical ratio  
$y=\mu_2^{1/(d-2}=( {L}/{\ell})^{1/(d-2)}$,
\begin{eqnarray}
\label{fin}
c_1 y^2+c_2 y+k=0,
\end{eqnarray}
where
%\begin{eqnarray}
%\label{fin1}
$
c_1=n_1^{2/(d-2)}\left( {V_{\kappa,d-3}}/{V_{\kappa,d-2}}\right)^{2/(d-2)}
$
,
$
c_2=-\left( {V_{\kappa,d-3}}/{V_{\kappa,d-2}}\right)^{1/(d-2)} ({n_2}/{n_1^{(d-3)/(d-2)}})\frac{1}{d-2}.
$
%\end{eqnarray}
For $\kappa=1$, this argument predicts the existence of a critical value of $r_h$ 
above which the black strings are stable ($i.e.$ no real
solutions of the equation (\ref{fin})).
However, it also gives two possible values for the ratio ${L}/{\ell}$ 
(which are in the same range with what we have found
numerically).
Moreover, this argument predicts an instability of the topological 
black strings for suitable values of the event horizon area, which
appear  not to occur.
This may be due to the fact that the SAdS black holes would fail 
to provide a suitable lowest order approximation
for the hypothetical AdS counterparts of the $\Lambda=0$ caged black holes.
 
Concerning more general solutions,  the study of the perturbative equations 
in second order for static $\Lambda=0$ black strings, revealed the appearance of a critical dimension,
above which the perturbative nonuniform black strings
are less massive than the marginally stable uniform black string
\cite{Sorkin:2004qq}.
It would therefore be interesting to solve the perturbative equations
to higher order also for AdS solutions. 
Also, the unstable black strings are part of a larger phase diagram.
Other classes of AdS black objects,
presenting the same structure of the
 conformal infinity but a different topology of the event horizon are likely to exist. 
 \\
 \\
{\bf Acknowledgments}\\
 YB gratefully acknowledges the Belgian FNRS for
financial support.  The work of ER was supported by
a fellowship from the Alexander von Humboldt Foundation
%%%%%%%%%%%%%%%%%%%%%%%%%%%%%%%%%%%%%%%%%%%%%%%%%%%%%%%%%%%%%%%%%%%%%%%%%%%%%%

\end{document}